\documentstyle[aps,times]{revtex}
\input epsf
\epsfverbosetrue

\begin{document}
\draft \sloppy \twocolumn[
\hsize\textwidth\columnwidth\hsize\csname
@twocolumnfalse\endcsname

\title{Shot noise of series quantum point contacts intercalating chaotic cavities}

\author{S.~Oberholzer$^{1}$\footnote{present address: Delft University
of Technology, PO Box 5046, 2600 GA Delft, The Netherlands},
E.~V.~Sukhorukov$^2$, C.~Strunk$^3$, and  C.~Sch\"onenberger$^1$}
\address{
$^{1}$University of Basel, Klingelbergstr.~82, 4056 Basel, Switzerland\\
$^{2}$University of Geneva, 24, quai Ernest Ansermet, 1211 Geneve 4, Switzerland\\
$^{3}$University of Regensburg, D-93040 Regensburg, Germany}

\date{\today}
\maketitle

\begin{abstract}
Shot noise of series quantum point contacts forming a sequence of
cavities in a two dimensional electron gas are studied
theoretically and experimentally. Noise in such a structure
originates from local scattering at the point contacts as well as
from chaotic motion of the electrons in the cavities. We found
that the measured shot noise is in reasonable agreement with our
theoretical prediction taking the cavity noise into account.
\end{abstract}
]

\section{Introduction}
Shot noise is a non-equilibrium type of electrical current noise
    directly resulting from random transfer of discrete
    charge quanta \cite{BlanterReview1999}.
    For Poissonian transfer of single electrons the spectral density
    of the current
    fluctuations is $S_{Poisson}= 2e|I|$.
    Correlations imposed by Fermionic statistics or Coulomb
    interaction may change shot noise from $S_{Poisson}$ which is
    expressed by the Fano factor $F$ defined as $F = S/S_{Poisson}$.
    This is for example the case for a one mode quantum wire with an intermediate
    barrier of transmission probability $T$ where the shot noise
    is suppressed by a Fano factor $F = 1-T$ below its full Poissonian
    value \cite{LesovikJEPT1989,Kumar96}.
    In various mesoscopic systems universal Fano
    factors have been found such as $F = 1/3$ in metallic diffusive wires
    \cite{Beenakker1992,HennyPRB1999} or $F = 1/4$ in chaotic cavities
    \cite{JalabertEPL1994,BlanterPRL2000,OberholzerPRL2001}.

In this article the shot noise of a series of quantum point
    contacts forming a sequence of cavities in a two-dimensional
    electron gas is studied theoretically and experimentally.
    The noise in such a structure originates from local
    scattering at the point contacts as well as from diffraction
    of the electronic wavefunction within the cavities.
    While the Fano factor $F$ of a single scatterer equals $1-T$,
    with $T$ its transmission probability,
    $F$ reaches $1/3$ in the limiting case of an
    infinite number of scatterers \cite{DeJongPRB1995}.
    Thus the case of a large number of point contacts in series
    models a diffusive wire with randomly placed
    impurities, for which
    the Fano factor is also $1/3$ \cite{Beenakker1992,HennyPRB1999}.
    The work presented here is devoted to this crossover
    from $F=1-T$ for a single scatterer to $F=1/3$
    in the diffusive regime.
    Experimentally, this
    can be investigated by measuring shot noise of several quantum point
    contacts (QPC) in series, which model the impurity scattering
    in a diffusive wire.

\section{Crossover from a single scatterer to the diffusive regime}
The shot noise of a sequence of $N$ planar tunnel barriers has
    been calculated by de Jong and Beenakker
    within a semiclassical description based on the
    Boltzmann-Langevin approach \cite{DeJongPRB1995}.
    For equal transmission probabilities $T_{i=1,\ldots,N}=T$
    the Fano factor is found to be
\begin{figure}[h]
\centering \epsfxsize=84 mm  \epsfbox{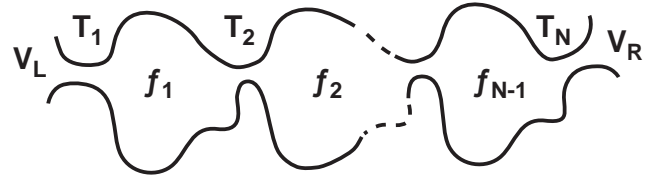}
\vspace{8 pt} \caption{Schematics of the considered system: $N$
    quantum point contacts forming a series of
    cavities. $f_{n=0.\ldots,N}$ denote
    the distribution functions of the electrons. In such a system shot
    noise arises due to quantum diffraction inside the cavity as
    well as to partitioning at the contacts.}
\label{barriers}
\end{figure}
\begin{equation}
    F = \frac{1}{3} \left(1+
    \frac{N(1-T)^2(2+T)-T^3}{[T+N(1-T)]^3} \right).
    \label{deJong}
\end{equation}
    According to this result $F$ indeed reaches 1/3 with
    increasing barrier number
    (\mbox{$N\rightarrow\infty$)} for \emph{any} value of the
    transparency $T\in [0,1]$ \cite{comment1}.
    In this description a
    diffusive conductor can be modeled as the continuum limit of a
    series of tunnel barriers.
    However, the calculation in Ref.~\cite{DeJongPRB1995}
    is only valid for
    one-dimensional transport since it neglects transverse motion of the
    electrons in between the barriers \cite{comment2}.
    It is therefore not appropriate
    to describe our physical system consisting of a series of quantum
    point contacts [Fig.~\ref{barriers}].
    Here, we consider the case that there are cavities between
    the barriers in which the electrons scatter chaotically leading
    to additional cavity noise \cite{JalabertEPL1994,BlanterPRL2000,VanLangenPRB1997}.

In general, the fluctuations in the total current through the
    system shown in Fig.~\ref{barriers}
    can be written as \cite{BlanterPRL2000,VanLangenPRB1997}
\begin{equation}
    \delta I = \delta I_{n}^S + G_{n}(\delta V_{n-1}-\delta V_{n}),
    \quad n=1,\ldots, N
    \label{currentfluctuations}
\end{equation}
    using the fact that the total current is conserved.
    $G_{n}=G_{0}\sum_{k}T_{kn}$ ($G_{0}\equiv 2e^2/h$) is
    the conductance of the $n$-th QPC and
    $\delta V_{n}$ the voltage fluctuations inside the $n$-th cavity.
    $\delta I_{n}^S$ are the 
    current fluctuations
    of a single QPC:
\begin{equation}
        \langle \delta I_{n}^{S}\delta I_{m}^{S}\rangle  =
    S_{n}\delta_{nm} 
    \label{single} \vspace{-1mm}
\end{equation}
    with \cite{LesovikJEPT1989}
\begin{eqnarray}
    S_{n} & = & 2G_{0}\sum_{k} \int_{0}^{\infty}
    dE\,\Bigl[ T_{kn} f_{n-1}(1-f_{n-1})
    \nonumber\\
    && + \, T_{kn} f_{n}(1-f_{n})\nonumber\\
    && + \, T_{kn}(1-T_{kn}) (f_{n-1}-f_{n})^2 \Bigr].
    \label{Lesovik2}
\end{eqnarray}
    Summing the square of Eq.~(\ref{currentfluctuations}) over $n$,
    while assuming equal
    conductances $G_{n}=G$, the total noise power follows as
\begin{equation}
    S \equiv \langle \delta I^2\rangle = \frac{1}{N^2}\sum_{n,m=1}^{N}
    \langle \delta I_{n}^{S}\delta I_{m}^{S}\rangle =
    \frac{1}{N^2}\sum_{n=1}^{N}S_{n},
    \label{totalnoise}
\end{equation}
    where we have assumed $\delta V_{0}=\delta V_{N}=0$, i.e. no
    fluctuations in the potential of the perfect metallic leads.

\subsection{Non-interacting electrons}
The distribution function $f_{n}$ of the $n$-th cavity follows
    from the conservation of numbers of
    electrons in each energy interval \cite{BlanterPRL2000}.
    For equal conductances of the $N$ point contacts $f_{n}$ is given by
\begin{equation}
    f_{n}(E)=\left(\frac{N-n}{N}\right)f_{L}(E)+
    \frac{n}{N}f_{R}(E),
    \label{ffcold}
\end{equation}
    with $f_{L}(E)=f_{F}(E,eV_{L},\theta)$ and $f_{R}(E)=
    f_{F}(E,eV_{R},\theta)$ the equilibrium Fermi function in the
    left and right reservoir, respectively. $\theta$ denotes the bath
    temperature.
    For simplicity only one propagating mode will be considered for the
    moment with the backscattering parameter
    $\mbox{$\cal{R}$}_{n}\equiv 1-T_{n}=\cal{R}$ assumed to
    be the same for all barriers.
    Substituting the distribution function $f_{n}$ from
    Eq.~(\ref{ffcold})
    into Eq.~(\ref{Lesovik2}), the total
    noise of $N$ point contacts in series follows from
    Eq.~(\ref{totalnoise}):
\begin{eqnarray}
    S & = & \frac{4G_{S}k_{B}\theta}{N^2} \biggl\{
        \frac{1}{3}\Bigl[(2N^2+1) \nonumber\\
       &&  + \, (N^2-1)\frac{eV}{2k_{B}\theta}
       \coth \left( \frac{eV}{2k_{B}\theta} \right)\Bigr]\nonumber\\
        &&  + \, \mbox{$\cal{R}$}\Bigl[\frac{eV}{2k_{B}\theta}
        \coth\left( \frac{eV}{2k_{B}\theta}
        \right)-1\Bigr]\biggr\}. \label{complicated}
\end{eqnarray}
    $G_{S}=G/N$ is the total conductance of the device.
    In the zero-temperature limit we obtain for the Fano factor
\begin{equation}
    F\equiv \frac{S}{2e|I|} = \left( \frac{1}{3}-\frac{1}{3N^2}+
    \frac{\cal{R}}{N^2}\right).
    \label{barriernoisecold}
\end{equation}
    With this expression for the Fano
    factor $F$, Eq.~(\ref{complicated}) can be rewritten in a
    more simple and transparent form:
\begin{eqnarray}
        S & = & S_{eq}\biggl[\frac{2}{3}+\frac{1}{3}-
    \left(\frac{1}{3}-\frac{1}{3N^2}+
    \frac{\mbox{$\cal{R}$}}{N^2}\right)\nonumber\\
    && \hspace{1cm} + \, \left(\frac{1}{3}-\frac{1}{3N^2}+
    \frac{\mbox{$\cal{R}$}}{N^2}\right)(\beta\coth \beta)
    \biggr] \nonumber\\
          & = & S_{eq}\Bigl[1+F(\beta \coth\beta-1)\Bigr],
\end{eqnarray}
    with $S_{eq}=4G_{S}k_{B}\theta$ and $\beta\equiv (eV/2k_{B}\theta)$.

\begin{figure}[h]
\centering  \epsfxsize=86 mm \epsfbox{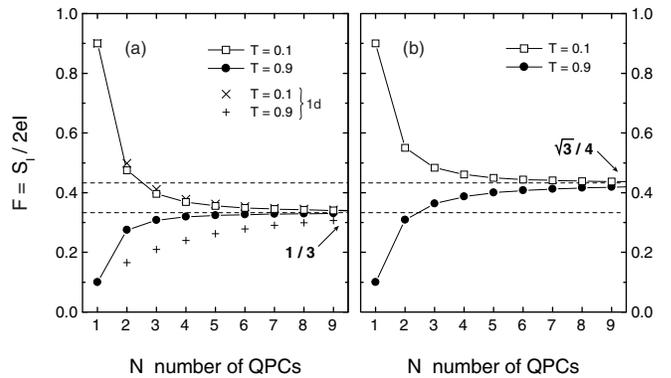}
\vspace{8 pt} \caption{Theoretical prediction for shot noise
normalized to the
    Poissonian limit $2e|I|$ as a
    function of the number of QPCs $N$ in series for cold (a) and
    hot electrons (b) at zero bath temperature.
    The crosses ($\times,+$)
    in the left plot correspond to a
    calculation of de~Jong and Beenakker [10]
    for one-dimensional
    transport, not including the noise arising from chaotic
    motion of the
    carriers in the cavities between the barriers.
    This is taken into account for Eq.~(\ref{barriernoisecold}).}
\label{theory}
\end{figure}

For a single QPC ($N=1$) $F$ equals the backscattering
    parameter $\mbox{$\cal{R}$}=1-T$ as expected.
    For $N=2$ a single cavity is
    separated from the leads by two QPCs and
    for ideal contacts (i.e. $\mbox{$\cal{R}$}=0$)
    the Fano factor is $1/4$.
    If the QPCs are in the tunneling regime ($\mbox{$\cal{R}$}\simeq 1$)
    the noise is dominated by
    the QPCs and the dynamics inside the cavity
    play no role.
    The Poissonian voltage noise of the two contacts adds
    up resulting in a Fano factor $1/2$.
    In the intermediate regime
    $F = \frac{1}{4}(1+\cal{R}).$
    Increasing the number
    of QPCs ($N\rightarrow\infty$) the Fano factor $F$
    reaches $1/3$ \emph{independent} of $\mbox{$\cal{R}$}=1-T$ as
    for the calculation in Ref.~\cite{DeJongPRB1995}.

In Fig.~\ref{theory}(a) the result of Eq.~(\ref{barriernoisecold})
    is compared to the result for one-dimensional tunnel-barriers
    \cite{DeJongPRB1995}.
    For point contacts with low transparencies ($T=0.1$) the results are
    very similar because in this case the noise is
    dominated by the contacts.
    But for high transparencies
    ($T=0.9$) the Fano factor including `cavity noise'
    [Eq.~(\ref{barriernoisecold})] increases much
    faster with the number of QPCs than the one-dimensional model
    of Ref.~\cite{DeJongPRB1995}.

\subsection{Interacting electrons}
\begin{figure}[h]
\centering \epsfxsize=80 mm \epsfbox{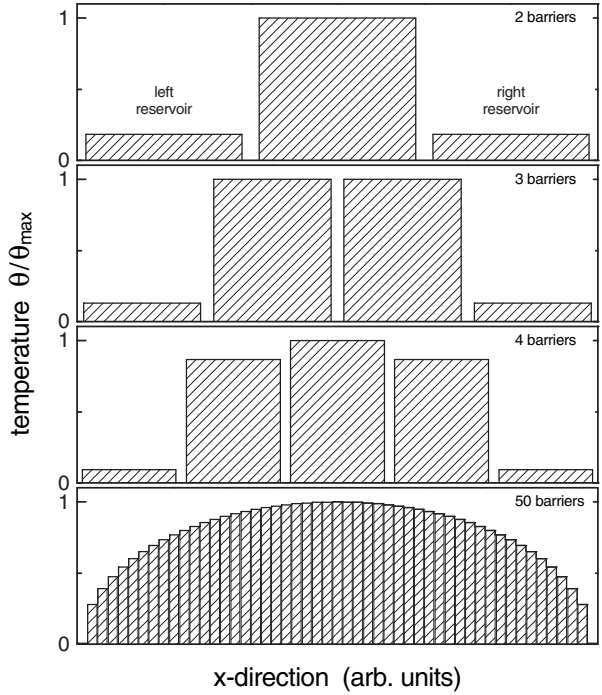}
\vspace{8 pt}
\caption{Temperature profiles for %
    2, 3, 4 and 50 barriers in series.
    In case of inelastic
    electron-electron scattering electrons
    thermalize within the cavities
    and the temperature drastically increases above the bath temperature
    leading to enhanced noise.
    For a large number of barriers the temperature profile has a
    parabolic shape as for a metallic diffusive
    wire [14].}
\label{tempprofile}
\end{figure}

In case of electron-electron interaction within the cavities the
    distribution function inside the $n$-th cavity $f_{n}(E,\theta)$
    equals a Fermi function $f_{F}(eV_{n},\theta_{n})$
    at an elevated electron \mbox{temperature $\theta_{n}$}
    [Fig.~\ref{tempprofile}].
    $V_{n}=V(1-n/N)$ is the potential in the $n$-th cavity
    with $V=V_{L}-V_{R}$ the potential difference between
    left and right reservoir. \index{heating}
    The electron temperature $\theta_{n}$ follows from the energy
    balance equation using the Wiedemann-Franz
    law \cite{SukhorukovPRB1999}:
\begin{equation}
        \theta_{n}^2 =
    \theta^2+\frac{3n(N-n)}{N^2}\left(\frac{eV}{\pi k_{B}}\right)^2.
        \label{eq:tttt}
\end{equation}
    Using Eq.~(\ref{Lesovik2}) and (\ref{totalnoise}) the noise
    in this case of hot electrons is given by
 \noindent %
\begin{eqnarray}
    S & = & \frac{2}{N^2}\sum_{n=1}^{N}
    \biggl\{Gk_{B}(\theta_{n}+\theta_{n-1})
    \nonumber \\
    && +\sum_{k}\int_{0}^{\infty} dE\, T_{kn}(1-T_{kn})
    \left[ f_{n-1}-f_{n}\right]^2 \biggr\}.
    \label{hhhhooootttt}
\end{eqnarray}
    The integral in Eq.~(\ref{hhhhooootttt}) cannot be
    calculated analytically for the general case.
    Numerical results for the Fano factor are
    shown in Fig.~\ref{theory}(b).
    Analytical expressions can be given for $T_{i=1,\ldots,N}=1$
\begin{equation}
    F=\frac{2\sqrt{3}}{\pi N^2} \sum_{n=1}^{N-1} \sqrt{n(N-n)}
    \label{hotballistic}
\end{equation}
    and for the tunneling regime ($T_{i=1,\ldots,N}\ll 1$) in the
    `diffusive limit' ($N\rightarrow \infty$)
\begin{equation}
    F = \frac{2\sqrt{3}}{\pi} \int_{0}^{1} dx \sqrt{x(1-x)}
    \label{hotdiffusive}
\end{equation}
    with $x=n/N \in [0,1]$.
    In both cases $F$ equals $\sqrt{3}/4$ in
    the  limit $N\rightarrow\infty$, in agreement with the
    result for a diffusive conductor with electron heating
    \cite{Nagaev95,Steinbach96}.

\section{The device}
Experimentally, a structure as described in
    Fig.~\ref{barriers} has been
    realized in a 2 DEG by a series of up to four QPCs across a
\begin{figure}[h]
\centering \epsfxsize=75 mm \epsfbox{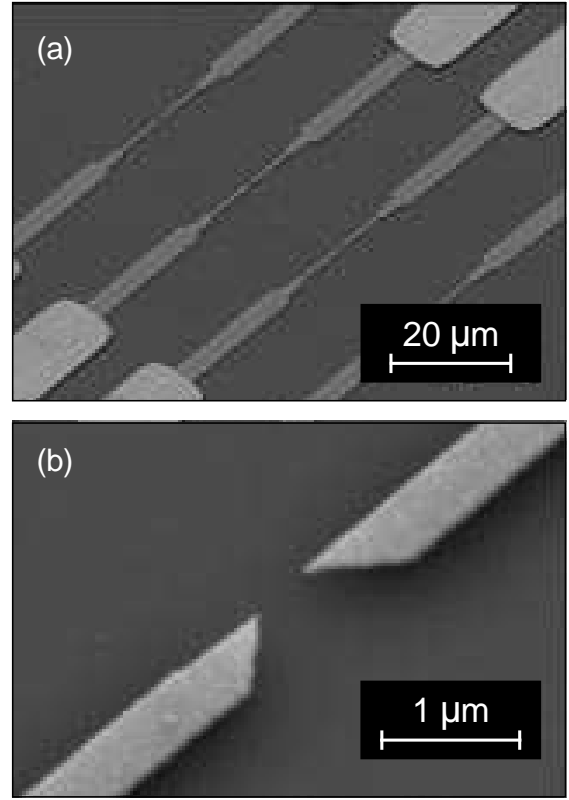}
\vspace{2 pt} \caption{Scanning electron
    microscope images
    of the measured device. (a) Four point contacts forming a sequence of
    large cavities. (b) The single point are defined with split gates.}
\label{barriersample}
\end{figure} \noindent %
    wet-chemical etched Hall-bar.
    The 2 DEG forms at the interface
    of a standard GaAs/Al$_{0.3}$Ga$_{0.7}$As-heterojunction
    80 nm below the surface.
    Its carrier density equals $2.7\cdot 10^{15}$ m$^{-2}$ and the
    mobility is 80 Vs/m$^{2}$.
    The different split gates are spaced by 20
    $\mu$m and the Hall-bar is 100 $\mu$m wide.
    Thus, the size of the cavity is rather large so that thermalization
    of the electrons can take place inside
    the cavity \cite{OberholzerPRL2001}.

Voltage noise has been measured as a function of current
    for one to three QPCs in series
    with fixed transmission probabilities.
    In the experiment,
    we used only up to three QPCs
    because one of the four did not show proper conduction
    quantization.
    Since the gates do not influence each other, the transmission of
    a single point contact can be determined independently by
    measuring its conductance while the others are kept completely
    open \cite{OberholzerPRL2001}.
    The spectral
    density of the voltage fluctuations are typically averaged over a
    frequency bandwith of 1 kHz at around 7 to 9 kHz.
    Experimental details can be found in \cite{HennyPRB1999,OberholzerPRL2001}

    \section{Results and discussion}
\begin{figure}[h]
\centering \vspace{3pt}\epsfxsize=80 mm
\epsfbox{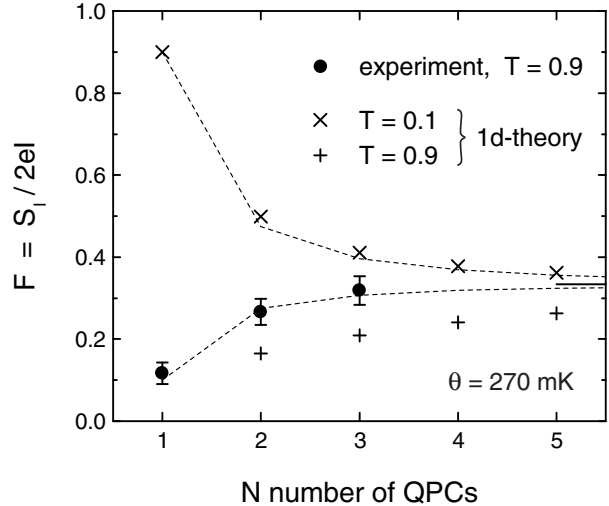} \vspace{8 pt}
\caption{The black points are %
    experimental data for one mode transmitted at
    the point contacts with
    probability \mbox{$T=0.9$.} The dashed line are theoretical
    predictions
    including cavity noise for non-interacting electrons
    [Eq.~(\ref{barriernoisecold})].
    The crosses correspond to the one-dimensional model
    of de Jong and Beenakker [10].}
\label{Fano}
\end{figure}

In Fig.~\ref{Fano} the Fano factor $F\equiv S/2e|I|$ extracted
    from the shot noise measurements
    is plotted as a function of the number of point contacts.
    The black dots correspond to experimental data
    obtained in the case that each single QPC is adjusted to a
    transmission probability $T\simeq 0.9$.
    The dashed lines are the prediction of
    Eq.~(\ref{barriernoisecold}) for $T=0.1$ (upper curve)
    and for $T=0.9$ (lower curve), whereas the crosses correspond to the
    one-dimensional calculation of Ref.~\cite{DeJongPRB1995}.
    For $N=1$ shot noise is strongly suppressed compared to
    $S_{Poisson}=2e|I|$, as expected for a single QPC with
    high transmission probability \cite{Kumar96}.
    When $N$ goes from 1 to 3 shot noise
    becomes larger.
    We find that the Fano factor increases faster with increasing
    number of the contacts $N$ than predicted by the
    one-dimensional theory.
    For $N=3$ it is already very close to the $1/3$ value of the
    diffusive regime. The measured data are consistent with our
    theoretical prediction that takes the cavities into account.
    This demonstrates that in the system depicted in Fig. 1 not only
    partition noise at the contacts contributes to the total noise but
    also additional cavity noise.
    Since the cavities are large compared
    to the mean free path of the electrons the electrons
    stay relatively long  in the cavity so that inelastical
    scattering should be present.
    However, the uncertainties of the experimental
    data due to small inequalities
    in the transparencies of the single QPCs and due
    to mode mixing do not allow to distinguish
    between the cold- and hot-electron regime in this experiment.

 \noindent %

    \section{Conclusions}
We studied the shot noise of a series of QPCs forming a sequence
    of cavities.
    Theoretical calculations along the lines of
    Ref.~\cite{BlanterPRL2000}
    show that the shot noise reaches $1/3$ and $\sqrt{3}/4$ of
    the Poissonian limit for cold and hot electrons, respectively, when
    the number of point contacts $N$ is increased to infinity.
    Noise measurements on a series of 1, 2 and 3 QPCs defined in a 2 DEG
    are in reasonable qualitative agreement with our calculation
    that takes the contribution from the cavities into account.

This work was supported by the Swiss National Science Foundation.



\begin{thebibliography}{99}
\bibitem{comment2}
    A series of planar tunnel barriers can be modeled by series
    QPCs in the integer Quantum Hall regime with no backsacttering in
    the cavities between the barriers.
\end{thebibliography}
\end{document}